\documentstyle[aps,prd,preprint,tighten,floats,epsf]{revtex}

\begin{document}
%

\preprint{VAND-TH-96-4}

\title{Interpolating the Stage of Exponential Expansion in
the Early Universe - a possible alternative with no reheating}

\author{Arjun Berera}

\address{
   Department of Physics and Astronomy,
   Vanderbilt University,
   Nashville, TN 37235, U.S.A.
   August 1996
}


\maketitle

\begin{abstract}
In the standard picture,
the inflationary universe is in a supercooled state which ends
with a short time, large scale reheating period, after which the universe
goes into a radiation dominated stage.  An alternative is proposed
here in which the radiation energy density smoothly decreases all
during an inflation-like stage and with no discontinuity enters
the subsequent radiation dominated stage. The scale factor is
calculated from standard Friedmann cosmology in the
presence of both radiation and vacuum energy density.
A large class of solutions confirm the above identified regime
of non-reheating
inflation-like behavior
for observationally consistent expansion factors and
not too large a drop in the radiation energy density. One dynamical
realization of such inflation without reheating is from warm inflation
type scenarios.  However the solutions found here are properties
of the Einstein equations with generality beyond slow-roll inflation
scenarios.  The solutions also can be continuously interpolated
from the non-reheating type behavior to the standard supercooled limit
of exponential expansion, thus giving all intermediate inflation-like
behavior between these two extremes.  The temperature of the universe
and the expansion factor are calculated for various cases.
Implications for
baryongenesis are discussed.  This non-reheating, inflation-like
regime also appears to have some natural features for a universe
that is between nearly flat and open.
\end{abstract}

\medskip

PACS numbers: 98.80.Cq, 05.40.+j

\medskip

In Press Physical Review D

\bigskip

hep-ph/9612239

\bigskip


\section{Introduction}
\label{sec.int}

In the original conception of inflation \cite{guth}, it was assumed
that the universe underwent isentropic expansion during
the  stage of rapid growth of the scale factor.
The entropy required to make the post-inflationary universe
consistent with observation was assumed to be generated
in a short-time
reheating period.  However, it is clear
that for a range of moderate thermodynamic conditions,
the cosmological horizon and flatness problems,
which are explained by inflation, require only
the kinematic property that the scale factor grows rapidly.  More
recently \cite{bf1} it was realized that these kinematic conditions could
still arise in the presence of a sustained radiation component 
during inflation.  Specifically, in \cite{bf1} it was shown that under
certain isothermal conditions inflation could still occur.
More so it was shown there, that within these limits, the initial seeds
of density perturbations could be dominantly of thermal
instead of quantum origin.  A realization of an isothermal or warm inflation
scenario, in the context of slow-roll
scalar field dynamics for
parametrically large dissipation, was shown in \cite{wi}
to be consistent with observational constraints
for the amplitude and expansion factor, without
requiring a ultra-flat Coleman-Weinberg potential, which
in order to form requires the coupling constant to be fine tuned.
Questions about the fundamental origin of large dissipation
are still left open.

The warm inflation scenario served as a explicit demonstration of an
otherwise true but ambiguous statement, that inflation can
occur in the presence of
a thermal component.  That this is true is
self-evident, as for example within the context of scalar field theory.
Here the
requirements for exponential expansion are 
\begin{equation}
\rho_v \gg \delta \rho_{\phi}
\end{equation}
and
\begin{equation}
\rho_v \gg  \rho_{\rm kinetic}
\end{equation}
with
\begin{equation}
\rho_{\rm kinetic} \equiv \frac{1}{2} {\dot  \phi}^2 + \rho_r .
\end{equation}
Here $\delta \rho_{\phi}$ is the energy density perturbation,
and $\rho_v$ and $\rho_r$ are the background vacuum and 
radiation energy densities
respectively.  Thus energetics alone does not prohibit the relation
\begin{equation}
\rho_v \gg \rho_r \gg \frac{1}{2} {\dot \phi}^2, \ \delta \rho_{\phi} .
\end{equation}
By itself this inequality
gives no indication on the extent that radiation can modify
the supercooled scenario. However,  the warm inflation scenario \cite{wi}
demonstrated that at least in the limit of near thermal equilibrium,
the effect is nontrivial.  By reexamining this scenario
solely in terms of energetics in \cite{ab2}, it became
evident that both supercooled and thermal slow-roll scenarios could be viewed
as limiting cases of a class of nonequilibrium
kinetic possibilities.  A preliminary step to a nonequilibrium study
is determining the possible kinematic behaviors of
the scale factor for a universe in a
mixed state of radiation and vacuum energy.  This is the first
motivation that leads us to examine the scale factor in
this paper.

In light of this, we find it useful to distinguish between the behavior of
the scale factor, which we consider kinematics, from the underlying
dynamics that induces this behavior.
The classification of scale factor behavior is considered
kinematic, because
it involves characterizing different solutions and different regimes
of a given solution, all arising from a particular
equation.  Besides inflation,
common amongst these are radiation dominated and matter
dominated behavior.    Originally inflation was associated with
an exponentially growing scale factor \cite{guth}. Subsequently
any form of accelerated expansion (${\ddot R}(t) > 0$) has
become associated with inflation.

Dynamics enters in determining the time evolution of the background
stress energy tensor, which is the driving source in the scale
factor equation.  In the context of dynamics, inflation
may or may not arise due to a phase transition.
In general, dynamics is stochastic, although the
degree of stochasticity may well be approximated by pure dynamics
or near-equilibrium statistical dynamics \footnote{The
role of stochasticity in cosmology has been emphasized
by the Maryland school. For a review please see \cite{hu1,hu2}}.  Inflation
scenarios realized in a supercooled regime are examples of the
former, whereas warm inflation scenarios \cite{wi}
are examples of the latter. 

The present most successful formulation of supercooled scenarios
is new inflation \cite{newi,brand}.  Although several variants of the
original scenario have been formulated (for a review please
see \cite{olive,brand3}), up to observation
the basic assumptions and mechanism are the same.
The new inflation assumptions are that dynamics can be
described by a suitable potential with a suitable order parameter,
known as the inflaton, and that evolution is governed by the
Lagrangian equations of motion.  The basic mechanism of new inflation
is slow-roll dynamics at supercooled temperatures. 

In the simplest form of new inflation, the inflaton
is a scalar field.
The conventional treatment of scalar field dynamics assumes that
it is pure vacuum energy dominated.  The various kinematic
outcomes are a result of specially chosen Lagrangians.
In most cases the Lagrangian is unmotivated from particle
phenomenology. Clear exceptions are the 
Coleman-Weinberg potential with an untuned coupling
constant, which
is motivated by grand unified theories \cite{ggmod}, and supersymmetric
potentials, although in the latter case, the choice of
the supersymmetric potential is again arbitrary, and in the
former case new inflation is inconsistent with observation.
Making one extension to the new inflation
picture, the behavior of the scale factor can also
be altered for any given potential when radiation energy is
present.  Out of pure kinematic interest, this effect has reason
to be examined. 

More so than just this reason,
one may also project to circumstances
sometime in the future when observational data will
allow determination of the optimal potential amongst the candidate choices
(for examples of recent attempts please see \cite{detpot1,detpot2}).
If one accepts the new inflation approximation that the relaxational dynamics
of the inflaton can be described by a potential, the next question 
is what is the microscopic origin of this so preferred potential.
If one were restricting to supercooled scenarios, one
argument is that the so preferred potential happens to be the one
that formed during the rapid quench at the onset of inflation.
Another argument is that this is a fundamental zero
temperature potential of an elementary field in the Lagrangian.
Since for supercooled new inflation scenarios, one of the
unanswered questions is that no potential that is suitable
for inflation has an already  known phenomenological origin,
the second argument is highly predictive.  Yet to substantiate
either claim, one would need to study the evolution
of the potential from its high temperature state during the quench.
This would lead to examining the interplay between radiation
and vacuum energy density at the onset of inflation.

Having appreciated this point, the time interval in
which this transition occurs
becomes important.  The short time regime is relevant
to supercooled new inflation scenarios and the extent to which
this interval can be extended is relevant to warm inflation
scenarios.  Thus, whether stated in the conventional sense of new inflation
or the extended sense of warm inflation, the out-of-equilibrium
evolution of the inflationary potential will require study, and 
as an initial step, the scale factor dynamics needs to be examined
in a mixed state of background vacuum and radiation energy density.
This additional connection to supercooled scenarios provides a second
motivation for this study.

To completely study the nonequilibrium dynamics, the problem divides
into two steps.
The first step is determining the
regimes in which accelerated expansion and pure inflation can occur
and characterizing the behavior of the scale factor in these regimes.
The second step is understanding within the
allowed regimes, the class of spectra
of primeval energy density perturbations.  
The first step is moderately model dependent and mainly involves
energetics and Friedman cosmology. The second step is a more
acute problem
of dynamics. Although we will only address the first step
in this paper, let us
make a few comments about the second step.

In general there is no unique
formulation of nonequilibrium dynamics
for almost any system. The first step in formulating
any approach requires understanding the scales in one's problem.
For inflation the simplest assumption is that there are two scales: a 
longtime, long-distance scale associated with vacuum energy
dynamics  and a single short-time, short-distance scale associated with
a random force component.   The Hubble time
during inflation, $1/H$, appropriately
separates the two regimes. For grand unified theory \cite{ggmod}, this
time interval is $1/H \approx 10^{-34} {\rm sec}$.

The assumption of a longtime scale for the evolution of the vacuum is based
on observation.  Otherwise inflation would not sustain itself sufficiently 
long nor would the energy release maintain smoothness.  Accepting this as an
empirical constraint, the relaxational dynamics of the inflaton's
order parameter justifiably could be described by a free-energy functional.
What the specific functional is requires dynamics. 
In the presence of a radiation component, the functional
need not have any similarity to a fundamental potential from the
underlying quantum field theory.  Furthermore, 
in grand unified theories as an example,
the characteristic
time scale of inflation, $1/H$, is about $10^{10}$ times faster than the
characteristic hadronic interaction scale ($1/\Lambda_{QCD}$), which
is a comparison scale where there is good empirical understanding
about matter. Thus at the inflation scale, familiar
concepts about matter and
from field theory about
near-thermal-equilibrium-motivated effective 
potentials also need not be appropriate.  

The problem here has similarities to certain phase separation problems
commonly known in association with
binary alloys, and the name synonymous with them, spinodal
decomposition, has been used before in new inflation cosmology
\footnote{For scalar field inflaton dynamics, the analogy
actually is to
spinodal decomposition for a nonconserved order parameter,
such as found in certain domain growth problems
\cite{allcah,bray}, whereas the binary alloy problem
involves a conserved order parameter}.
The similarity in both cases is that the system is being cooled
faster than its characteristic response time to equilibrate.
Of course if this analogy is meant to be complete, the cooled
system should still be at a nonnegligible temperature, since at
least in the binary alloy problem, the relaxational dynamics is driven by
short ranged thermally excited fluctuations.  The analogy
to spinodal decomposition not only gives a nice guiding
picture, but it
also has a type of consoling appeal, which covers for our ignorance
about matter, much less quantum field theory, under such
extreme conditions, since at least in the context of alloys, the problem is
considered sufficiently complex to make
phenomenological modeling of the nonequilibrium potential
an accepted practice. If viewed in the same way,
the several scalar inflationary potentials that have
been suggested could be interpreted as the cosmologist's attempt at
nonequilibrium phenomenology.

\subsection{Hypothesis}
\label{sec1a}

Although the reasons given above well motivate examination
of the scale factor,  I will now describe an alternative to
the standard inflationary universe scenario.
Consider the following possibility which will be demonstrated in the sequel.
It should be easy to convince oneself that a radiation energy density
$\rho_r(t)$ of
say one part in ten thousand to the vacuum energy density
$\rho_v(t)$, probably
should not alter too much the inflation-like behavior of the scale factor.
However when looked upon in terms of the temperature $T_r$ of
the radiation energy density, this implies that $T_r$ is only
an order of magnitude below the scale of the vacuum energy density.
If such a state for the radiation energy density could be maintained
by the mutual effects of constant vacuum energy  decay and a steadily
decreasing acceleration of the scale factor, it could be possible
for an inflation-like stage to smoothly enter into a radiation dominated
stage without any discontinuities in $\rho_r(t)$.  This possibility
was suggestive from formulating the warm inflation
scenario \cite{wi}.

In this paper evidence is presented for inflation-like trajectories
of the scale factor which solve the horizon and flatness
problems, but for which the radiation
energy density monotonically enters the post-inflation
radiation dominated stage with in particular no 
intermediate reheating stage.  This is a regime
in between the radiation
dominated and inflation regimes, which has features similar to
both a big-bang-like explosion and an inflation-like expansion.

The paper is organized as follows. In the next section the
problem is formulated, general solutions are given in section 3,
special examples are given in section 4 and finally
the conclusion is in section 5.  From the class of solutions
that we find for the scale factor, supercooled expansion,
which we call supercooled inflation, is a limiting case.
This is a kinematic identification.  A particular and
most noteworthy
dynamic realization of
supercooled inflation scenarios is the class of
new inflation scenarios.
Supercooled inflation has associated with it also a range
of power law \cite{pli}, quasi-exponential \cite{chainf}
and exponential \cite{newi,brand} behavior for the scale
factor. However these varied behavior arise from the
specifics of the particular Lagrangian that is being considered.
In what we will examine, for any
given Lagrangian, a large range of behavior may
still arise, depending on the radiation energy density content.

There is one other problem that the present work may help
clarify.  We will discuss it briefly here.
However it gets into the realm of field theory dynamics, which
this paper will mainly avoid.
The most notable shortcoming of new inflation is in
explaining small scale energy density inhomogeneities 
\cite{amp,brand,brand2,kolb,peebles}.
The problem is sometimes referred to as the amplitude fluctuation
problem \cite{brand2}.  The warm inflation scenario in \cite{wi}
is a solution to this problem.  However our formulation
there did not detail a time history for
an inflation-like state with radiation.  The present work does
and in fact was its starting motivation.
However due to the generality of the solutions given here, it
appears better to consider warm inflation as a particular
dynamic realization within the big-bang-like inflation
regime.


Our equations can also be examined for the initial stage of
entering into the rapid expansion state, but we will not study that
here.

\section{Formulation}
\label{sec2}

We are interested in the scale factor from some short time
after the initial singularity, when quantum gravitational
effects become negligible.  We assume space is
homogeneous and isotropic, and restrict ourselves to
Friedman cosmology with the Robertson-Walker metric
\begin{equation}
ds^2 = dt^2 - R^2(t)\left[\frac{dr^2}{1-kr^2}
+r^2d\theta^2+r^2 \sin^2\theta d\phi^2 \right].
\end{equation}
For notational convenience, the origin
of cosmic time is defined as the beginning of our
treatment.

Let us start with the standard equations of Friedman cosmology
\cite{kolb,weinberg} for the scale factor $R(t)$
in the presence of vacuum energy density $\rho_v(t)$ and
radiation energy density $\rho_r(t)$.  The equations of state
which relate the energy density $\rho$ to the pressure $p$
are
\begin{equation}
p_v(t) = -\rho_v (t)
\label{eqsv}
\end{equation}
\begin{equation}
p_r(t) = \frac{1}{3} \rho_r (t) .
\label{eqsr}
\end{equation}

In Friedman cosmology the ten Einstein equations
$G_{\mu \nu}= 8 \pi G T_{\mu \nu}$ reduce to two independent
ones, which are from the time-time component, also known as
Friedmann's equation
\begin{equation}
\frac{\dot{R}^2}{R^2}+\frac{k}{R^2}=\frac{8 \pi G}{3} \rho ,
\label{fried}
\end{equation}
and from any of the three diagonal space-space components, all
of which give
\begin{equation}
2 \frac{\ddot{R}}{R}+\frac{{\dot R}^2}{R^2}+\frac{k}{R^2}=
-8 \pi G p .
\end{equation}
For our purposes it is preferable to use two other equations obtained
from these, the scale factor equation
\begin{equation}
\frac{\ddot{R}}{R} = \frac{8 \pi G}{3} 
\left[\rho_v(t) - \rho_r(t)\right] ,
\label{scalefac}
\end{equation}
and the stress energy conservation equation
\begin{equation}
\dot{\rho}_r(t) = -4 \rho_r(t) \frac{\dot{R}(t)}{R(t)} - 
\dot{\rho}_v(t) ,
\label{econs}
\end{equation}
where we have used the equations of state eqs. (\ref{eqsv}) 
and (\ref{eqsr}).  We aim to solve for $R(t)$
and $\rho_r(t)$ in eqs. (\ref{scalefac}) and
(\ref{econs}), for a prescribed
$\rho_v(t)$, for $t>0$, and with arbitrary initial conditions
for $R(t)$, $\dot{R}(t)$, and $\rho_r(t)$ up to
the constraints
\begin{equation}
R(0) > 0
\label{con1}
\end{equation}
\begin{equation}
\dot{R}(0) > 0
\label{con2}
\end{equation}
and
\begin{equation}
\rho_r(0) > 0.
\label{con3}
\end{equation}

By taking the sum and difference of
Friedman's equation eq. (\ref{fried}) and the scale factor equation
eq. (\ref{scalefac}), the vacuum and radiation energy densities
can be separately expressed in terms of 
the scale factor as \cite{krzysztof}
\begin{equation}
\rho_v(t) = \frac{3}{16 \pi G} \left[ \frac{\ddot{R}}{R}
+\frac{\dot{R}^2}{R^2} + \frac{k}{R^2} \right]
\label{rhovsep}
\end{equation}
\begin{equation}
\rho_r(t) = \frac{3}{16 \pi G} \left[ -\frac{\ddot{R}}{R}
+\frac{\dot{R}^2}{R^2} + \frac{k}{R^2} \right].
\label{rhorsep}
\end{equation}
For an arbitrary test vacuum function $\rho_v(t)$, one can use
eq. (\ref{rhovsep}) to solve for $R(t)$.

We make the substitution
\begin{equation}
s(t) = R^2(t).
\label{s2}
\end{equation}
Eq. (\ref{rhovsep}) then becomes the inhomogeneous wave equation
with time dependent frequency
\begin{equation}
\ddot{s} - \frac{32 \pi G}{3} \rho_v(t) s = -2k.
\label{waveeq}
\end{equation}
This equation has been widely studied \cite{crc,kamke}.
Again we are interested in the solutions to eq. (\ref{waveeq})
for $t>0$ with arbitrary initial conditions for
$s(t)$ up to the constraints from eqs. (\ref{con1})-(\ref{con3})
and (\ref{s2}) which imply
\begin{equation}
s(0) > 0
\label{sg0}
\end{equation}
and
\begin{equation}
\dot{s}(0) > 0.
\label{dsg0}
\end{equation}

\section{Solution}
\label{sec3}

In this section solutions are obtained for the scale factor from
eqs. (\ref{scalefac})  and (\ref{econs}) for a large class
of vacuum energy decay functions.  Even before getting this specific,
there are two general features, one at short and one at long time,
which are recurring themes to the existence of the big-bang-like
inflation regime.
At long time, if $\rho_v(t)$ goes to
zero sufficiently fast, from eq. (\ref{waveeq}) and eq. (\ref{s2})
one can see that
$ R(t \rightarrow \infty) \sim t^{1/2} $, thus tending to a radiation
dominated behavior.  At short time, for an initially radiation
dominated universe,
\begin{equation}
\rho_r(t\sim 0) \gg \rho_v (t \sim 0) ,
\label{rrggrv}
\end{equation}
eqs. (\ref{scalefac}) and (\ref{econs}) imply
$R(t \sim 0) \sim (a +bt)^{1/2}$.  Alternatively, this can be seen
from Meissner's separation eqs. (\ref{rhovsep}) and (\ref{rhorsep}),
since eq. (\ref{rrggrv}) implies from eqs. (\ref{rhovsep}) and 
(\ref{rhorsep}) that 
\begin{equation}
\dot{s}^2(t \sim 0) \gg \ddot{s}(t \sim 0).
\label{sdggsdd}
\end{equation}
Taylor expanding $s(t)$ about the origin
as $s=s_0+ s_1 t + s_2 t^2/2 + \cdots s_n t^n/n! + \cdots +$,
eq. (\ref{sdggsdd}) implies $s_1^2 \gg s_2$. Using
this and eq. (\ref{waveeq}), one can study the initial condition
dependence of entering the inflation-like stage, but we will
not pursue that here.

Let us now turn to specific solutions.
Although this paper is focused on
the kinematic possibilities for the scale factor, independent of
justification from any specific field theory, we will motivate
a class of vacuum decay functions from a general class of
scalar field dynamics. In fact as we will show below, in the limit
of strong dissipation, this motivation can be partly justified. 

We consider stochastic evolution for the inflaton governed
by the Langevin-like equation
\begin{equation}
\ddot{\phi}(t) + \left[\Gamma + 3 \frac{{\dot R}(t)}{R(t)}\right] 
{\dot \phi}(t)
+V'(\phi (t)) = \eta(t),
\label{rateeq}
\end{equation}
where $\eta(t)$ is a random force function with 
vanishing ensemble averaged expectation
value 
\begin{equation}
\langle \eta(t) \rangle = 0.
\end{equation}
The effect of the inflaton's interaction with radiation is represented
by the dissipative constant $\Gamma$ and the random force function
$\eta (t)$.  In a simple model for the radiation system and in the limit
of pure inflation (${\dot R}/R=$const.) this equation was obtained
from quantum field theory in \cite{ab2}.

We are interested in the limit of strong dissipation
\begin{equation}
\Gamma \gg \frac{{\dot R}}{R}
\label{gbigh}
\end{equation}
and the slow-roll regime
\begin{equation}
\Gamma | \dot{\phi} | \gg | \ddot{\phi} |.
\label{gpdggpdd}
\end{equation}
For our present purposes, the ensemble averaged equation of motion
is all that we need. Thus in the above specified limits, eq.
(\ref{rateeq}) becomes
\begin{equation}
\frac{d \phi}{dt} = - \frac{1}{\Gamma} \frac{d V(\phi)}{d \phi}.
\label{slowroll}
\end{equation}
Let us consider potentials of the form
\begin{equation}
V(\phi) = \lambda M^{4-n} (M-\phi)^n
\label{approxv}
\end{equation}
in the region 
\begin{equation}
0 < \phi < M
\end{equation}
where $\lambda$ is dimensionless.
For inflation driven dynamically at the grand unified scale
$M \sim M_{GUT} \approx 10^{14}$ GeV.

Globally all of the potentials in eq. (\ref{approxv})
are improper for slow-roll inflation
scenarios, since they fail to represent symmetry breaking.  However
our present interest is the behavior of the scale factor
for a large class of slow-roll conditions.  In this sense
eq. (\ref{approxv}) represents a class of local approximating
potentials from which an arbitrary potential can be piecewise
constructed.  Thus it is also not a concern that such potentials
have no minima for odd n and are nonanalytic for noninteger n
when $\phi = M$.

In fact near the global minima, where the vacuum energy goes to zero,
quadratic ($n=2$) dependence would be the normal expectation
for any generic free energy functional.  This case is not only of special
physical interest but is also mathematically a little different. We
will differentiate this case of $n=2$ from all others and refer to
it as the quadratic limit.

To keep our discussion explicit, we will express the results that follow 
in the context of the slow-roll inflation scenario.  However it should
be noted that the solutions for the scale factor given below carry
a relevance beyond the slow-roll scenario. Let us briefly recall the
slow-roll scenario. In the standard setting of the slow-roll
transition, the inflaton starts near the origin and
is making its decent to the symmetry broken minima at $\phi=M$.

At the origin of cosmic time we will assume
the slow-roll transition begins with
\begin{equation}
\phi (0) = \epsilon M
\label{phi0}
\end{equation}
and $\epsilon \ll 1$.  With these initial conditions, the solutions
of eq. (\ref{slowroll}) for potentials in eq. (\ref{approxv})
are for $n=2$
\begin{equation}
\phi(t) = M \left[ 1- 
\exp \left( -\frac{B_2}{2} (t-t_{0_2}) \right) \right],
\label{phin2}
\end{equation}
and for $n \neq 2$
\begin{equation}
\phi(t) = M \left[1 - \left(B_n (t + t_{0_n}) 
\right)^{\frac{1}{2-n}} \right]
\label{phinn2}
\end{equation}
with
\begin{equation}
B_2 \equiv \frac{4 \lambda M^2}{\Gamma}
\label{B2}
\end{equation}
and
\begin{equation}
B_n \equiv \frac{n(n-2) \lambda M^{2}}{\Gamma}.
\label{Bn2}
\end{equation}
Here $t_{0_2}$ and $t_{0_n}$ are suitably 
adjusted to satisfy eq. (\ref{phi0}).
Equating the potential to the vacuum energy density
\begin{equation}
\rho_v(t) = V (\phi (t) )
\end{equation}
implies for $n = 2$
\begin{equation}
\rho_v (t) = \lambda M^4 \exp(-B_2 (t-t_{0_2}))
\label{vacn2}
\end{equation}
and for $n \neq 2$
\begin{equation}
\rho_v(t) = \lambda M^{4}
\left[ B_n ( t + t_{0_n}) \right]^{\frac{n}{2-n}} .
\label{vacnn2}
\end{equation}
Substituting the above in eq. (\ref{waveeq}) and solving the homogeneous
(flat space) equation, we find for the scale factor from eq. (\ref{s2})
for $n=2$
\begin{equation}
R = \sqrt{C_1 I_0(z_2(t)) + C_2 K_0(z_2(t))}
\label{soln2}
\end{equation}
and in the $n \neq 2$ case for all but $n=4$
\begin{equation}
R = \left[ B_n (t + t_{0_n}) \right]^{\frac{1}{4}}
\sqrt{C_1 I_{\frac{2-n}{4-n}}(z_n(t)) + 
C_2 K_{\frac{2-n}{4-n}}(z_n(t))},
\label{solnn2}
\end{equation}
where
\begin{equation}
z_2(t) \equiv \frac{4}{B_2} H_2 
\exp \left( \frac{-B_2}{2} t \right)
\label{z2n2}
\end{equation}
\begin{equation}
z_n(t) \equiv 
\frac{4(2-n)H_n t_{0_n}}{(4-n)}
\left( \frac{t}{t_{0_n}}  + 1 \right)^{\frac{4-n}{4-2n}}
\end{equation}
with
\begin{equation}
H_2 \equiv \sqrt{ \frac{8 \pi G \lambda M^4}{3} }
\exp \left( \frac{B_2 t_{0_2}}{2} \right)
\label{hub2}
\end{equation}
\begin{equation}
H_n \equiv \sqrt{ \frac{8 \pi G \lambda M^{4}}{3} }
(B_n t_{0_n})^{\frac{n}{2(2-n)}}.
\label{hubn2}
\end{equation}
In eqs. (\ref{hub2}) and (\ref{hubn2}) we have identified
the Hubble parameter at $t=0$ based on the definition
\begin{equation}
H \equiv \sqrt{\frac{8 \pi G \rho_v(0)}{3}}
\label{hubdef}
\end{equation}
for the respective vacuum energy densities in eqs. (\ref{vacn2})
and (\ref{vacnn2}).
In the Appendix we have listed properties 
of Modified Bessel functions that
will be useful to us.  Finally from the $n \neq 2 $ cases for $n=4$
the solution is
\begin{equation}
R=|B_4 ( t + t_{0_4} )|^{1/4}
\sqrt{C_1 |B_4 ( t + t_{0_4} )|^{\mu} +
C_2 |B_4 ( t + t_{0_4} )|^{-\mu}}
\label{soln4}
\end{equation}
where
\begin{equation}
\mu = \frac{1}{2} \left(1+ 
\frac{128 \pi G \lambda M^4}{3B_4^2} \right)^{1/2}.
\end{equation}

The inhomogeneous wave equation in eq.
(\ref{waveeq}),
which is for curved space $k \neq 0$, can be solved from the above
solutions for the homogeneous equation by familiar methods
\cite{crc,kamke}.  Irrespective of the slow-roll scenario,
the results eqs. (\ref{soln2}) and (\ref{solnn2}) are valid
for any scenario that motivates vacuum decay behavior as in
eqs. (\ref{vacn2}) and (\ref{vacnn2}). Likewise for
other types of vacuum decay functions, eq. (\ref{waveeq}) can be solved
\footnote{One case
is during reheating in supercooled scenarios.  For this,
the vacuum decay function in eq. (\ref{waveeq}) should have
the approximate time dependence $e^{-|\Gamma |t} (1+\cos Bt)$
with $B \sim M \gg H$. These types of
equations are treated in \cite{crc}.
The $|\Gamma|=0$ case is the Mathieu equation. The solutions
of these equations describe
the scale factor behavior during those stages of reheating
when the equation of state eq. (\ref{eqsv}) is valid for the
inflaton.
Another nontrivial aspect of scale factor
behavior in supercooled scenarios is at the
beginning where 
initial condition dependence on pre-inflationary
radiation energy density can be studied.  For this, the
solutions eq. (\ref{solnn2}) for $n \sim 0$ are useful.}.

In the next section,  the solutions eqs.
(\ref{soln2}) and (\ref{solnn2})
will be studied through specific examples.
Here some of their general features will be noted.
The quadratic limit is examined first.  The growing mode
as $t \rightarrow \infty$ in eq. (\ref{soln2}) from
eqs. (\ref{api00}) and (\ref{apk00}) is 
$K_0(z_2(t))$ with
\begin{equation}
R(t \rightarrow \infty) \sim t^{1/2},
\end{equation}
thus asymptotically exhibiting radiation dominated behavior.

Inflation-like expansion at intermediate time is also governed
by $K_0(z_2(t))$. From eq. (\ref{apk00}) a large expansion factor
of $e^N$ with $N \geq 50$ will require 
\begin{equation}
\frac{2H_2}{B_2} \sim N.
\label{estn2}
\end{equation}
This is like what one would expect, since
the vacuum energy density must
decay sufficiently slowly relative to the expansion time
for the scale factor, in order to be the driving source
for inflation-like behavior in the Einstein equations.

In order to establish the dominance of the $K_0(z_2(t))$ term to the
$I_0(z_2(t))$ term in eq. (\ref{soln2}), what remains is to
show that there is no way for the initial conditions
to force $C_1$ to be exponentially large relative to
$C_2$.  Treating $4H_2/B_2 \gg 1$ and using eqs. (\ref{apini})
and (\ref{apkni}), this follows from the constraints eqs. (\ref{sg0})
and (\ref{dsg0}).
Eq. (\ref{dsg0}) could be satisfied for $C_1$ exponentially large 
but negative relative to $C_2$,
but then eq. (\ref{sg0}) would not be satisfied. 
Having established the dominance of the $K_0(z_2(t))$
mode, let us estimate the expansion
factor for a single $n=2$ section of potential eq. (\ref{approxv})
with $\epsilon = 0$ in eq. (\ref{phi0}) so that $t_{0_2}=0$
in eq. (\ref{phin2}).  We find for the asymptotic behavior
\begin{equation}
\frac{R(t \rightarrow \infty)}{R(0)} \sim
\frac{(B_2t/2)^{1/2} e^{2H_2/B_2}}{(\pi B_2/8H_2)^{1/4}}
\end{equation}
so an expansion factor of $e^{2H_2/B_2}$.

Away from the quadratic limit ($n \neq 2$) from eqs. (\ref{solnn2}),
(\ref{apini}), (\ref{apkni}) we see that for
\begin{equation}
\frac{4-n}{2-n} > 0
\end{equation}
the solution will grow exponentially at large time, thus never
asymptotes into a radiation dominated behavior. This corresponds
to a vacuum decay function that decays slower than $1/t^2$
at large times in eq. (\ref{waveeq}).
Radiation dominated behavior at large time is attained for
\begin{equation}
\frac{4-n}{2-n} < 0
\end{equation}
which implies potentials in eq. (\ref{approxv}) with
\begin{equation}
2 < n < 4
\label{n24}
\end{equation}
or vacuum decay functions in eq. (\ref{waveeq}) that decay faster
than $1/t^2$. Finally for $n=4$, which corresponds to
a vacuum decay falling-off exactly as $1/t^2$, $R(t)$ has the same
power law behavior throughout, with a growth bounded
from below by $t^{1/2}$. As such, this case is not useful
for our present purpose.  This implies that the only symmetric
potential about the symmetry broken point, $\phi=M$,
that leads to radiation dominated and not inflation-like
asymptotic behavior is the quadratic case $n=2$. As an aside,
note that the $n=4$ case 
is interesting since on
either side are solutions with two very different types
of asymptotic behavior \footnote{It is also an interesting
coincidence that $n=4$
separates renormalizable and nonrenormalizable scalar
quantum field theory, with the nonrenormalizable side, $n > 4$,
corresponding to the observationally inconsistent
non-radiation dominated asymptotic scale factor behavior.
Furthermore the $n=4$ case neither
asymptotes to radiation dominated behavior nor is believed to be
nonperturbatively a nontrivial quantum field theory \cite{wilson}.
Of course for the inflaton, since it is coupled to gravity, the
whole theory is always nonrenormalizable in any case.}.

Returning to the cases in eq. (\ref{n24}),
the growing mode in eq. (\ref{solnn2})
is $K_{\frac{2-n}{4-n}}(z_n(t))$. 
Let us estimate the expansion factor for a single
$n \neq 2$ sector of the potential eq. (\ref{approxv})
in the range eq. (\ref{n24}) for $\epsilon = 0$ in
eq. (\ref{phi0}) so that 
\begin{equation}
B_n t_{0_n}= 1
\label{t0n2}
\end{equation}
in eq. (\ref{phinn2}).
The arguments are the same as above for the $n=2$ case with the
final result
\begin{equation}
\frac{R(t \rightarrow \infty)}{R(0)} \sim
(B_n t)^{1/2}
\exp \left[\frac{2(n-2)H_n}{(4-n) B_n} \right] .
\label{appnnn2}
\end{equation}

Before closing this section, one additional qualifying statement is
needed about the solutions eqs. (\ref{soln2}) and (\ref{solnn2})
if the vacuum decay functions eqs (\ref{vacn2}) and 
(\ref{vacnn2}) are
obtained from slow-roll scalar field dynamics.  Recall that the energy
density and pressure of the zero mode of the scalar field
are
\begin{equation}
\rho_{\phi} = \frac{1}{2} {\dot \phi}^2 + V(\phi)
\end{equation}
\begin{equation}
p_{\phi} = \frac{1}{2} {\dot \phi}^2 - V(\phi).
\end{equation}
Therefore, the equation of state eq. (\ref{eqsv}) is valid
in the limit that the potential energy dominates
the kinetic energy
\begin{equation}
\frac{1}{2} {\dot \phi}^2  \ll V(\phi).
\label{kinsup}
\end{equation}
One must check that this kinetic energy suppression
condition is always valid.  

For this, first recall that the exact
equation of motion for the inflaton in the limit
eq. (\ref{gbigh}) is the second order
equation
\begin{equation}
{\ddot \phi} + \Gamma {\dot \phi} + V'(\phi) = 0.
\end{equation}
For the quadratic case $n=2$, this equation can be exactly solved
and it can be verified that the slow-roll condition
eq. (\ref{gpdggpdd}) and the kinetic energy suppression condition
eq. (\ref{kinsup}) are both valid for all $t>0$ provided
\begin{equation}
\frac{\lambda M^2}{\Gamma^2} \ll 0.
\label{gbigml}
\end{equation}
In addition, if $\rho_v(0)$ is required to be large in eqs. (\ref{vacn2})
and (\ref{vacnn2}), so that $\lambda$ can not be made tiny,
eq. (\ref{gbigml}) implies
\begin{equation}
\Gamma \gg M.
\label{gbigm}
\end{equation}
This is the large dissipative regime required for warm inflation
\cite{wi}.

For the cases $2 < n < 4$, to verify eq. (\ref{kinsup}),
first it will be shown that the solutions eq. ({\ref{phinn2})
are consistent with the slow-roll condition eq. ({\ref{gpdggpdd})
for all $t > 0$.  Next a direct verification will be made
that the solutions eq. (\ref{phinn2}) respect the condition
eq. (\ref{kinsup}).  Addressing step one, it is observed from
eq. (\ref{phinn2}) that for the entire range $2<n<4$,
${\ddot \phi}(t)$ vanishes faster than ${\dot \phi}(t)$
as $t \rightarrow \infty$.  Thus eq. (\ref{gpdggpdd}) is
satisfied under the same parametric restrictions 
as in the $n=2$ case, eqs. (\ref{gbigml})
and (\ref{gbigm}).  Proceeding to the second step, it can be
verified from the slow-roll approximate solutions
eq. (\ref{phinn2}) that ${\dot \phi}^2 (t)$ vanishes faster than
$V(\phi (t))$ as $t \rightarrow \infty$.
Thus in the regime eq. (\ref{gbigm}), eq. (\ref{kinsup})
is satisfied for all $t>0$ so that eq. ({\ref{eqsv})
is always valid.

To summarize, it has been verified that the equation of
state eq. (\ref{eqsv}) is valid for the scalar field for all
$t>0$ and in the entire range $2 \leq n < 4$, when in
the strong dissipative regime eq. (\ref{gbigm}).
This type of slow-roll motion is analogous to an over-damped
oscillator \cite{ab2}.  Note that confirming the validity of
eq. (\ref{eqsv}) for all $t>0$ is more than needed, since
in any case $\rho_r(t)$ overtakes $\rho_v(t)$ at some 
much earlier stage.

The results presented in this section now demonstrate the existence of
inflation-like scale factor trajectories which smoothly
go into a radiation dominated behavior without a discontinuous
reheating stage.

\section{Examples}
\label{sec4}

In this section we will examine some specific examples from
the solutions for the scale factor in eqs. (\ref{soln2})
and (\ref{solnn2}).  In these examples we will see how
the radiation energy density eventually overtakes the vacuum
energy density with no
discontinuities, and in the processes the universe smoothly
goes from an inflation-like to a radiation dominated stage.
We will also study the magnitude of decrease in the radiation
energy density, thus the temperature of the universe,
from before to after the inflation-like stage.  In supercooled
scenarios, the post-inflation temperature is referred to as the
reheating temperature, but here it is better to call it
the initial temperature after inflation, $T_{AI}$.
In particular the inflation-like stage is defined as the time period when
the scale factor has positive acceleration
\begin{equation}
\ddot{R}(t) > 0,
\end{equation} 
with the time ``just before'', $t_{BI}$, and 
``just after'', $t_{AI}$,
inflation
being defined as the endpoints of the accelerated
expansion interval

In subsection 4b we will examine a particular $n \neq 2$ case
from eq. (\ref{solnn2}) which can be fully expressed with
simple analytic functions.  Then in subsection 4c we will
examine the quadratic limit.  For this study, we will
first convert to a set of dimensionless quantities.

\subsection{The Dimensionless Theory}
\label{sec4a}

We will work with the dimensionless quantities defined as
\begin{equation}
a(\tau) \equiv \frac{R(\tau)}{R(\tau_{BI})}
\label{adef}
\end{equation}
\begin{equation}
b(\tau) \equiv \frac{\rho_v(\tau)}{\rho_v(\tau_{BI})}
\end{equation}
\begin{equation}
c(\tau) \equiv \frac{\rho_r(\tau)}{\rho_v(\tau_{BI})}
\end{equation}
where dimensionless time
\begin{equation}
\tau \equiv H t,
\end{equation}
$\tau_{BI}$ is the
time when accelerated expansion begins, and
$H$ is defined in eq. (\ref{hubdef}) except with the vacuum energy
density evaluated at $\tau_{BI}$, $\rho_v(\tau_{BI})$.
Defining $s_a(\tau) \equiv a^2(\tau)$, the Meissner separation,
eqs. (\ref{rhovsep}) and (\ref{rhorsep}),
in terms of $s_a(t)$ and the rescaled quantities
is
\begin{equation}
b(\tau) = \frac{1}{4s_a(\tau)}
\frac{d^2s_a(\tau)}{d\tau^2}
+ \frac{k}{2H^2 s_a(\tau)}
\label{msb}
\end{equation}
and
\begin{equation}
c(\tau) = -\frac{1}{4s_a(\tau)}
\frac{d^2s_a(\tau)}{d\tau^2}
+\frac{1}{4s^2_a(\tau)}
\left( \frac{ds_a(\tau)}{d\tau} \right)^2
+ \frac{k}{2H^2 s_a(\tau)}.
\label{msc}
\end{equation}

The radiation energy density will be related to a temperature
measure by the Stefan-Boltzmann radiation law
\begin{equation}
\rho_r(\tau) \sim T^4 (\tau).
\end{equation}
This law need not hold under far from equilibrium conditions,
but we will nevertheless refer to $T(\tau)$ as the temperature
of the universe at time $\tau$.
We will study the temperature of the universe in terms
of the ratio 
\begin{equation}
\alpha (\tau) \equiv \frac{T(\tau)}{T(\tau_{BI})}.
\label{temprat}
\end{equation}

The field theory quantities will also be rescaled.  The 
natural scale for them is $M$ not $H$ and in general
these two scales are different.  This scale disparity is
an inherent feature of scalar field slow-roll dynamics.
A primitive source of the dilemmas encountered in
slow-roll scenarios is its two scale nature.  The natural
time scale in the field theory to release the vacuum energy
$1/M$, in general differs from the characteristic cosmological
expansion time $1/H$. In grand unified theories this disparity
works against theoretical preference, since
\begin{equation}
1/M \ll 1/H.
\end{equation}
Had this inequality been reversed, it would have been parametrically
satisfying and perhaps a strong argument for theoretical consistency
between cosmology and particle physics.  However since this is
not the case, it either means slow-roll dynamics is wrong,
field theory dynamics for
inflation at the grand unified scale is wrong, grand unified
theory is incomplete or wrong or that the physics 
needs further elaboration, perhaps
from nonequilibrium methods.
We will not address the dynamical problem here, but
it is worthwhile to keep track of the scale disparity.
Thus we will rescale everything with respect to $H$, but for
quantities where $M$ is the natural scale, the rescaling will
include the additional factor
\begin{equation}
\beta \equiv \frac{M}{H} .
\end{equation}
The field theory quantities are rescaled as
\begin{equation}
\Gamma \equiv \gamma \beta H
\end{equation}
and
\begin{equation}
\phi \equiv \sigma \beta H.
\end{equation}
 
\subsection{n=8/3}
\label{sec4b}

Let us consider the case $n=8/3$ from the $n \neq 2$
class of potentials in eq. (\ref{approxv}) which in rescaled parameters
is
\begin{equation}
V(\sigma) = \lambda M^4 (1-\sigma)^{8/3}.
\end{equation}
Solving the slow-roll equation of motion
\begin{equation}
\frac{d \sigma}{d\tau} = \frac{3}{2} \kappa_{8/3} (1-\sigma)^{5/3} ,
\end{equation}
where using eqs. (\ref{Bn2})
\begin{equation}
\kappa_{8/3} \equiv \frac{B_{8/3}}{H_{8/3}}
= \frac{16 \lambda \beta}{9 \gamma}
\label{kap83}
\end{equation}
and with the initial condition
\begin{equation}
\sigma(0)=0 ,
\end{equation}
we find
\begin{equation}
\sigma(\tau)=1 -
\frac{1}{ (\kappa_{8/3} \tau + 1 )^{3/2} }.
\end{equation}
This implies that the rescaled vacuum energy density is
\begin{equation}
b(\tau)= \frac{1}{(\kappa_{8/3} \tau +1)^4}.
\label{vacb}
\end{equation}
Solving the homogeneous wave equation eq. (\ref{waveeq}) using
eq. (\ref{vacb}) we find
\begin{equation}
a(\tau) = \sqrt{(\kappa_{8/3} \tau +1) \left[
C_1 \exp \left( \frac{2 \tau}{(\kappa_{8/3} \tau + 1)} \right)+
C_2 
\exp \left( \frac{-2 \tau}{(\kappa_{8/3} \tau + 1)} \right) \right] }.
\label{asol}
\end{equation}
Here $\tau_{BI}$, $C_1$ and $C_2$ are determined by the initial
radiation energy density $r$,
\begin{equation}
\frac{c(0)}{b(0)} =r,
\label{cond1}
\end{equation}
the defining relation for $\tau_{BI}$,
\begin{equation}
b(\tau_{BI}) = c(\tau_{BI}),
\label{cond2}
\end{equation}
and from eqs. (\ref{adef}) which implies
\begin{equation}
a(\tau_{BI}) = 1.
\label{cond3}
\end{equation}
From eqs. (\ref{msb}) and (\ref{msc}) and restricting to
flat space, explicitly the first two
conditions above imply from eq. (\ref{cond1})
\begin{equation}
\frac{1}{4 s_a^2(0)} \left(\frac{ds_a(0)}{d\tau}\right)^2 =r+1
\label{cond11}
\end{equation}
and from eq. (\ref{cond2})
\begin{equation}
\frac{1}{8 s^2_a(\tau_{BI})} 
\left(\frac{ds_a(\tau_{BI})}{d\tau}\right)^2=b(\tau_{BI}).
\label{cond22}
\end{equation}
In eq. (\ref{cond22}) $s_a^2(\tau_{BI})=1$, but we retain it explicitly
since the same equality holds at $\tau_{AI}$, and we will use it
below to determine $\tau_{AI}$.

Let us verify the various general features discussed in earlier
sections for this specific example.  At large time, by inspection
of eq. (\ref{asol}) one finds
\begin{equation}
a(\tau \rightarrow \infty) \sim \tau^{1/2},
\end{equation}
which verifies an asymptotic radiation dominated behavior.
Turning to the growth of the scale factor, to obtain
an exponentially large one from eq. (\ref{asol})
at large time, the only way is from the first term (the growing
mode) and only if $1/\kappa_{8/3} \gg 1$.  The constraints
eqs. (\ref{sg0}) and (\ref{dsg0}) imply $C_1 > |C_2|$,
so that the growing mode can not be suppressed due to an
exponentially small coefficient $C_1$ relative $C_2$.
To leave no ambiguity,  let us focus on one case amongst a
large class which are all about the same for what we
want to study.  For simplicity, in the case we will consider,
arrange the initial conditions so that inflation begins at the
origin of cosmic time $\tau_{BI} = 0$.  One can confirm
from eqs. (\ref{cond3}) and (\ref{cond11}) that $C_1$ and
$C_2$ are about the same order of magnitude, so that at large
time, only the growing mode need be retained.  For
completeness we find
\begin{equation}
C_1 = \frac{1}{2} \left(
1 + \sqrt{2} - \frac{\kappa_{8/3}}{2} \right)
\label{c1}
\end{equation}
and
\begin{equation}
C_2 = \frac{1}{2} \left(
1 - \sqrt{2} + \frac{\kappa_{8/3}}{2} \right).
\label{c2}
\end{equation}

From the growing mode in eq. (\ref{asol}) we see that
to obtain $N_e$ e-folds of expansion requires
\begin{equation}
\frac{1}{\kappa_{8/3}} = N_e,
\label{necon}
\end{equation}
which agrees with our general $n \neq 2$ approximation formula
eq. (\ref{appnnn2}).
In grand unified theories, for the Coleman-Weinberg
potential with an untuned coupling
constant, one has $\beta \sim 10^4$ and $\lambda \sim 1$
so that from eqs. (\ref{kap83}) and (\ref{necon}) this implies
\begin{equation}
\gamma \sim N_e 10^4,
\end{equation}
which for $N_e \sim 50$ implies $\gamma \sim 10^6$.
This gives an overviewed explanation for the
large dissipative constant found in the warm inflation
scenario of \cite{wi}. However the estimate here for
$\gamma$ is a little higher, because in the actual
scenario, the finite temperature Coleman-Weinberg
potential has a smaller curvature. In this
simplified discussion, this means $\beta$ is smaller.

We see once again that the largeness of $\beta$ makes a seemingly
undesired appearance in the dynamics. Whereas in supercooled
scenarios it forces a fine tuning of the coupling constant,
here it forces the dissipative constant to be larger
than one would naively want. It remains a theoretical question
whether such a large dissipative constant has an explanation.
However up to the parametric level, the warm inflation
approach still has a hope of salvaging the simplest
grand unified theory motivated Coleman-Weinberg model
for inflaton dynamics.  In new inflation, this possibility
is not even parametrically satisfying, since
fine tuning the coupling constant also implies that the
vacuum energy, and so the Hubble constant during
inflation, drops substantially.

Turning to the radiation energy density next, our interest
is to compute from the ratio in eq. (\ref{temprat}) 
$\alpha (\tau_{AI})$.  To keep an explicit example
in mind, we again take $\tau_{BI}=0$ with $C_1$
and $C_2$ as given in eqs. (\ref{c1}) and (\ref{c2}).
$\tau_{AI}$ is determined by a solution once again to
$b(\tau) = c(\tau)$, so for a second solution
to eq. (\ref{cond22}) for $\tau_{AI} > \tau_{BI} = 0$.
One clearly expects a second intersection, since $b(\tau)$,
which goes as $(\kappa_{8/3} \tau +1)^{-4}$, is falling-off
faster than $c(\tau)$, which at large time goes as
$(\kappa_{8/3} \tau +1)^{-2}$, and after the first intersection
at $\tau_{BI}$, $b(\tau) > c(\tau)$.
To determine $\tau_{AI}$, only the growing mode of
$s_a(\tau)$ is retained from $(ds(\tau)/d\tau)^2$,
to give the relation
\begin{equation}
\frac{\kappa^2_{8/3}}{4} (\kappa_{8/3} \tau_{AI} +1)^2
+\kappa_{8/3} (\kappa_{8/3} \tau_{AI} +1)-1 =0.
\end{equation}
Solving for $\tau_{AI}$ with the constraint $\tau_{AI} > 0$ we find
\begin{equation}
\tau_{AI} = \frac{1}{\kappa_{8/3}}
\left[ \frac{2}{\kappa_{8/3}}(\sqrt{2}-1)-1 \right] \approx 
\frac{2(\sqrt{2}-1)}{\kappa^2_{8/3}},
\end{equation}
so that from eq. (\ref{msc})
\begin{equation}
c(\tau_{AI}) \approx \frac{\kappa^4_{8/3}}
{16(\sqrt{2}-1)^4}
\end{equation}
and
\begin{equation}
\alpha(\tau_{AI}) = 
\left(\frac{c(\tau_{AI})}{c(\tau_{BI})}\right)^{0.25}
\approx 1.2 \kappa_{8/3}.
\end{equation}
Recalling  from eq. (\ref{necon}) that $N_e$ e-folds of expansion requires
$\kappa_{8/3} = 1/N_e$, if $N_e \sim 50 - 70$ we find that
the temperature drops by a factor $1/50 - 1/70$
from the beginning to the end of inflation with the
duration of inflation being 
$\tau_{AI} - \tau_{BI} \sim N_e^2 \sim 2500-4900$.

To summarize, in this subsection
we have presented an example that can be verified
by inspection in which the scale factor expands sufficiently
and then smoothly tends towards a radiation dominated behavior at large
time. In the course of this, the temperature of the universe
during the inflation-like stage drops between 1-2 orders
of magnitude. 

\subsection{The Quadratic Limit}
\label{sec4c}

Let us next examine the quadratic limit with the vacuum decay
function
\begin{equation}
b(\tau) = \exp (-\kappa_2 \tau).
\label{vacdec}
\end{equation}
The plots are in figure 1 for the inverse Hubble
parameter
\begin{equation}
\frac{1}{H} \equiv \frac{a(\tau)}{da(\tau)/d\tau}
\end{equation}
and in figure 2 for the temperature ratio $\alpha(\tau)$ defined
in eq. (\ref{temprat}).  In both figures the solid (dashed)
curve is for the vacuum decay coefficient $\kappa_2 =0.03(0.04)$.
The calculations were done by a numerical integration of
the coupled scale factor equation (\ref{scalefac}) 
and stress-energy conservation
equation (\ref{econs}), with the vacuum decay function
in eq. (\ref{vacdec}). As a cross check, at each iteration
the so computed scale factor was substituted into the 
left-hand-side of Friedmann's equation (\ref{fried}),
with the resulting energy density found from the right-hand-side
compared with that from the numerical integration.
This also means the results cross check 
separately for $b(\tau)$ and
$c(\tau)$ from (\ref{msb}) and (\ref{msc}), but since this
requires $\ddot{a}(\tau)$, it is not a more stringent
test.

For both cases in Fig. (1), the inverse Hubble parameter starts out
flat, which is characteristic of inflation-like behavior.
It then veers up at a time of order $\tau \sim 33 (25)$ in the solid
(dashed) case  and finally tends to a slope of 2 at large times,
thus becoming a radiation dominated universe on schedule.
Inflation begins in the solid (dashed) case at 
$\tau_{BI}=0.073 (0.074)$ and ends at $\tau_{AI}= 323 (228)$,
so that the duration of inflation is 
$\tau_{AI} - \tau_{BI} \approx 323 (228)$.
During the inflationary period, the scale factor expands rapidly
with total e-folds $N_e = 67 (51)$, which agrees with
estimates from our approximation formula eq. (\ref{estn2}),
and the temperature drops
by a factor $\alpha(\tau_{AI}) = 11 (10)  $.

The numerical results for $\tau_{AI}$ and $\alpha(\tau_{AI})$
also can be cross checked to approximate analytic expressions,
Solving eq. (\ref{cond22}) for the second solution at
$\tau_{AI} > \tau_{BI}$, noting from the Appendix
that $K_0(z_2(\tau))$ dominates in the solution
eq. (\ref{soln2}) at long time, and using $dK_0(z)/dz = -K_1(z)$,
one finds for any $\kappa_2$ the general relation
\begin{equation}
\frac{K_1 \left( z_2(\tau_{AI}) \right)}
{K_0 \left( z_2(\tau_{AI}) \right) }
= \sqrt{2}.
\end{equation}
From \cite{abst} one finds that this is 
satisfied for $z_2 \approx 1.05$,
so that from eq. (\ref{z2n2}) we obtain the approximate
formula
\begin{equation}
\tau_{AI} \approx - \frac{2}{\kappa_2} 
\ln \left(\frac{1.05 \kappa_2}{4} \right).
\label{tai}
\end{equation}
Assuming $\tau_{BI} \ll 1$ so that $c(\tau_{BI}) \approx 1$,
as for the numerical cases presented above, and using
eq. (\ref{tai}) we obtain
\begin{equation}
\alpha(\tau_{AI}) \approx \frac{1}{2} \sqrt{1.05 \kappa_2}.
\label{altai}
\end{equation}
One can verify that the approximation formulas eqs. (\ref{tai})
and (\ref{altai}) reproduce the results for $\tau_{AI}$
and $\alpha(\tau_{AI})$ respectively that were quoted
above from the numerical calculations.  Similar approximation
formulas
can also be obtained for the $n \neq 2$ cases.

In Fig. (2) observe the initial steep drop in $\alpha(\tau)$
for $\tau < 1$.  There is a very short initial transient
period in which the initial radiation energy density
stabilizes, followed by a steady state stage.  For both cases
in Fig. (2), we started with a radiation energy density
$c(0) = 1.5$.  Thus initially
the first term on the right-hand-side of equation (\ref{econs})
(the "sink term")  rapidly depletes $\rho_r(t)$
(equivalently $c(\tau)$ in the rescaled theory)
until an
approximate balance is reached by the second term (the "source term"),
after which steady state is reached.  
The initial conditions on the radiation energy density
have a mild effect on the longtime behavior.
For example,
increasing $c(0)$ by a factor of 500 has less than a
one percent effect on $N_e$.  Without the source term, which
arises from vacuum energy depletion,
all the radiation energy would rapidly red-shift away,
as in supercooled scenarios.

\section{Conclusion}
\label{sec5}

There are two possibly concerning or possibly predictive
outcomes of scenarios occurring entirely in the 
big-bang-like regime
We believe they are general features of such scenarios,
although we do not have proof.
Firstly to attain an observationally consistent expansion factor,
it does not appear possible for the post inflation temperature $T_{AI}$
to be the same order of magnitude as that just before inflation
$T_{BI}$. In supercooled scenarios this is referred
to as a perfect reheating and can be achieved by adjusting
the decay width, which controls the reheating time period,
to be sufficiently large \cite{albrecht,kolb}.
In big-bang-like scenarios, for observationally sufficient
expansion, we
generally find that $T_{AI}$ is
at least one order of magnitude below $T_{BI}$.
Thus for inflationary dynamics at the grand unified scale one expects
$T_{AI} \sim (0.1 - 0.01)M_{GUT}$.  In the context of
grand unified theory, this implies the X-boson, with
$M_X \sim M_{GUT}$, would not participate in post-inflation
baryongenesis,
although the lighter Higgs boson still could \cite{kolb2}.
Moreover the picture is further altered since in 
the big-bang-like class of inflation
scenarios, there would be no violent
discontinuities in $\rho_r(t)$ at the end of 
the inflation-like stage.  This implies baryongenesis could commence within
the inflation-like stage and smoothly carry-on afterwards.
One can also consider long sustained big-bang-like inflation
scenarios, in which the temperature drops
by a few orders of magnitude during the inflation-like stage.
For such scenarios, studies of baryongenesis from
out-of-equilibrium decay processes at temperatures well
below $M_{\rm GUT}$ may be useful \cite{hall}.
As a final complementary note to this concern
pertaining to baryongenesis,
the lower temperature condition implies
magnetic monopole suppression works effectively.

The second point of concern for
big-bang-like inflation scenarios with not too large a drop in the
temperature during the inflation-like stage is that they
generally appear to have not very
large upper bounds on the expansion factor with e-folds
$N^{max}_e \sim 1000$ but $N^{max}_e \sim 100$ being typical.
Since observation indicates $N_e > 50-70$, this is still
acceptable. For comparison, in the solution of the
Coleman-Weinberg model
in new inflation, it is found that $N_e \sim 10^7$
\cite{brand}.
In general
new inflation models are reported to
predict very large e-folds $N_e$ \cite{kolb}.  As one optimistic
interpretation about the small e-fold constraint for
scenarios within the big-bang-like inflation regime, 
this is preferred if the universe is
between nearly flat and open \cite{ellis}.

In this paper we have shown that for a large range of vacuum energy
density decay trajectories, an early universe initially in a radiation
dominated stage can enter an inflation-like stage and finally
enter back into a radiation dominated stage with the radiation
energy density suffering no sharp alterations during this motion
and with a post inflationary temperature within a range consistent
with observation and theory.  This regime differs from the standard
inflation regime where the radiation energy density quickly vanishes
at the onset of inflation and then is quickly regenerated at the end of
inflation in a short time reheating era.  We reemphasize that the
solutions we have found are properties of the Einstein equations,
independent of quantum field theory. We also reiterate that in the
presence of nonnegligible radiation, one need not be restricted to
familiar near equilibrium quantum field theory methods in searching
for dynamical models.  This is not to preclude conventional
treatments.  In fact, despite our emphasis on the kinematic properties
of the final answer and its model independent origin, one should
note that we motivated all our results from the conventional
dynamical picture.

One cannot say without further investigation what the relevance
of the present results are. It has been established by this study that
sufficiently-rapid expansion behavior is more general than only
that found in the inflation regime.
As with any
generalization, there is always a danger that it is nothing more than a
mathematical novelty offering no new physical insight.  In the present
case, this does not appear to be a correct statement.  Firstly, 
in light of the new list of options, there
seems no special reasons that favor the supercooled limit to
any of the other possibilities demonstrated here. Also if the naturality
principle carries the interpretation that any possibility
not otherwise ruled out by observation nor theoretical common
sense is a candidate solution, then again the present generalization
has substance.  Finally in conjunction with warm inflation \cite{wi},
a suggestive solution to the amplitude fluctuation problem presents
itself.  However a dynamical explanation for large dissipation,
which is needed for that scenario, requires investigation.
On the other hand, cosmic string formation
\cite{cs1,cs2,cs3,olive},
which is typically considered a
post-inflation mechanism for large scale structure
or beginning at the end stage of a supercooled
scenario \cite{shavil}, could be a possible mechanism
within a large period of a
big-bang-like inflation scenario.

The most interesting result from this study is the finding
of an inflation-like regime of scale factor behavior that asymptotes
to the radiation dominated regime without a reheating
stage.  However returning to the introductory comments,
the solutions presented here have applicability also to
the class of supercooled inflation scenarios. In addition
by decreasing any of the coefficients $B$ in
eqs. (\ref{B2}) and (\ref{Bn2}), one can
smoothly interpolate from the big-bang-like stage
of inflation to the supercooled stage of exponential
expansion.
There are many possibilities suggested by our results from mild but
long sustained accelerated expansion to the standard exponential inflation.
With any of these, due to the presence of radiation, the dynamic
explanation may require the
range from familiar methods of finite temperature
quantum field theory to a full nonequilibrium statistical mechanical
treatment \footnote{Some considerations for reheating,
such as in \cite{hol}, may be useful also here.  In addition
further
examination could be made of the possible
nonequilibrium potentials (or free energy functionals) that
can form.  An example from spinodal decomposition is \cite{langer},
which starts from a master equation and attempts to deduce
the free energy functional.  This treatment was for a conserved
order parameter, whereas for inflaton dynamics a similar
treatment is needed for a nonconserved order parameter.
Other approaches which may be useful are in \cite{gleiser}. Finally
the works in \cite{fintemp} offer guidance in formulating the nonequilibrium
problem in an expanding universe.}.
Both further theoretical research and experimental
information is needed to narrow the possibilities.  

There are two extreme points of view that one might adopt.  One is
that the very early universe mostly wants to be radiation
dominated but just sneaks into a inflation-like phase
for a little while.  The other is that the very early
universe has trouble containing radiation energy just after
the initial singularity, so copiously inflates
until some heating or reheating mechanism finally stabilizes the
radiation energy.  The former is the mildest modification
of pre-inflation era thinking and the latter reflects
present thinking.  
For now, experiment and theory
do not indicate a strong preference for either viewpoint.
However it is a useful exercise to view the problem
from both extremes, since from either end
the other looks like a remote limiting case.  This in
our opinion is symptomatic of a misunderstanding about
the radiation energy content in the very early universe.
As such, we believe 
theories that make no
presumptions about the radiation content all during the early
universe better represent the
present status of experimental information about this time period.
Thus, allowing for any of the possible scale factor behaviors
derived here appears a more realistic starting point to further study.

\section*{Acknowledgments}
I Thank Krzysztof Meissner for his contribution and encouragement,
Tom Kephart for a careful reading of the final manuscript, and
the following for helpful discussions:
Misha Eides, Jorge Pullin, Li-Zhi Fang,
Bei-lok Hu, Zygmunt Lalak, Benjamin P. Lee,
Raul Abramo, Jane Charlton, Pablo Laguna, Lee Smolin and Abbay Ashtekar.
This work was support by the U. S. Department of Energy.
The initial  work began at the Pennsylvania State University
also under D.O.E. support.
 
\appendix
\section{}
\label{app1}

This appendix contains some properties of the Modified Bessel
functions $K_{\nu}(z)$ and $I_{\nu}(z)$, that are useful
for the results in the text.  At small z, the asymptotic behavior
for $\nu=0$ are
\begin{equation}
I_0(|z| \rightarrow 0) \sim 1
\label{api00}
\end{equation}
\begin{equation}
K_0(|z| \rightarrow 0) \sim - \ln |z|
\label{apk00}
\end{equation}
and for $\nu \neq 0$
\begin{equation}
I_{\nu}(|z| \rightarrow 0) \sim 
\frac{\left(\frac{1}{2} |z|\right)^{\nu}}{\Gamma(\nu +1)}
\label{apin0}
\end{equation}
\begin{equation}
K_{|\nu |}(|z| \rightarrow 0) \sim
\frac{1}{2} \Gamma(\nu)
\left(\frac{1}{2} |z|\right)^{-| \nu |},
\label{apkn0}
\end{equation}
where eq. (\ref{apin0}) is valid for all $\nu$ except
$\nu \neq -1, -2, \cdots$.

At large $z$ for all $\nu$
\begin{equation}
I_{\nu}(|z| \rightarrow \infty) 
\sim \frac{\exp(|z|)}{\sqrt{2 \pi |z|}}
\label{apini}
\end{equation}
\begin{equation}
K_{\nu}(|z| \rightarrow \infty) 
\sim \sqrt{\frac{\pi}{2|z|}} \exp(-|z|).
\label{apkni}
\end{equation}
Finally 
recall that
\begin{equation}
K_{\nu}(z) = K_{-\nu}(z),
\end{equation}
and for a negative argument
\begin{equation}
I_{\nu}(-|z|) = e^{i \nu \pi} I_{\nu}(|z|)
\label{apim}
\end{equation}
\begin{equation}
K_{\nu}(-|z|) = e^{-i \nu \pi} K_{\nu}(|z|)
- i \pi I_{\nu} (|z|).
\label{apkm}
\end{equation}

\eject
FIGURE CAPTIONS

figure 1: The inverse Hubble parameter for a quadratic slow-roll
potential with vacuum decay coefficients for the
solid (dashed) cases $\kappa_2 = 0.03 (0.04)$.
The initial conditions are $c(0)=1.5$.

figure 2: The ratio $\alpha(\tau)$ of the universe's
temperature at cosmic time $\tau$ to that at the beginning
of inflation $\tau_{BI}$ for the same cases as in
Fig. 1. For the solid (dashed) curve, the 
inflation-like stage begins at $\tau_{BI}=0.073 (0.074)$
and ends at $\tau_{AI} = 323 (228)$. In both cases the temperature
of the universe drops by about a factor 10, with e-folds
67 (51).


\begin{references}
\bibitem{guth} A. H. Guth, Phys. Rev {\bf D23}, 347 (1981).
\bibitem{bf1} A. Berera and L. Z. Fang, Phys. Rev. Lett. {\bf 74},
1912 (1995).
\bibitem{wi} A. Berera,  Phys. Rev. Lett. {\bf 75},
3218 (1995).
\bibitem{ab2} A. Berera, Phys. Rev. {\bf D54}, 2519 (1996).
\bibitem{newi} A. Albrecht and P. J. Steinhardt, Phys. Rev. Lett.
{\bf 48}, 1220 (1982); A. Linde, Phys. Lett. {\bf 108B}, 389 (1982).
\bibitem{brand} J.M. Bardeen, P.J. Steinhardt, \& M.S. Turner, 
Phys. Rev. {\bf D28}, 679 (1983);
R. Brandenberger and R. Kahn, Phys. Rev. {\bf D29}, 2172 (1984).
\bibitem{olive} K. A. Olive, Phys. Rep. {\bf 190}, 307 (1990).
\bibitem{brand3} R. H. Brandenberger, Rev. Mod. Phys. {\bf 57},
1 (1985).
\bibitem{ggmod} H. Georgi and S. L. Glashow, Phys. Rev. Lett.
{\bf 32}, 438 (1974).
\bibitem{detpot1} E. W. Kolb, A. Abney, E. J. Copeland,
A. R. Liddle, and J. E. Lidsey, FERMILAB-CONF-94-189-A
(astro-ph/9407021).
\bibitem{detpot2} R. L. Davis, H. M. Hodges, G. F. Smoot, P. J.
Steinhardt, and M. S. Turner, Phys. Rev. Lett. {\bf 69}, 1856 (1992).
\bibitem{pli} F. Lucchin and S. Mataresse, Phys. Rev. {\bf D32},
1316 (1985).
\bibitem{chainf} A. Linde, Phys. Lett. {\bf 129B}, 177 (1983).
\bibitem{amp} A. H. Guth and S. -Y. Pi, Phys. Rev. Lett.
{\bf 49}, 1110 (1982); S. Hawking, Phys. Lett. {\bf 115B},
295 (1982); A. A. Starobinskii, Phys. Lett. {\bf 117B}, 175 (1982).
\bibitem{brand2} R. H. Brandenberger, Nucl. Phys. {\bf B245},
328 (1984).
\bibitem{kolb}
E.W.Kolb and M.S. Turner,
{\it The Early Universe}, (Addison-Wesley, New York, 1990).
\bibitem{peebles} P.J.E.Peebles, {\it Principles of Physical
Cosmology}, (Princeton, 1993).
\bibitem{weinberg} S. Weinberg, {\it Gravitation and Cosmology:
Principles and Applications of the General Theory of
Relativity}, (John Wiley and Sons, New York, 1972).
\bibitem{krzysztof} This separation was pointed out 
to me by Krzysztof A. Meissner. 
\bibitem{crc} A. D. Polyanin and V. F. Zaitsev,
{\it Exact Solutions for Ordinary Differential Equations},
(CRC Press, New York, 1995).
\bibitem{kamke} E. Kamke,  
{\it  Differentialgleichungen L\"{o}sungsmethoden
und L\"{o}sungen}, (Chelsea, New York, 1948).
\bibitem{abst} M. Abramowitz and I. A. Stegun,
{\it Handbook of Mathematical Functions},
(Dover, New York, 1972).
\bibitem{albrecht} A. Albrecht, P. J. Steinhardt, M. S. Turner
and F. Wilczek, Phys. Rev. Lett. {\bf 48}, 1437 (1982);
A. D. Dolgov and A. D. Linde, Phys. Lett. {\bf B116},
329 (1982); L. F. Abbott, E. Farhi, and
M. B. Wise, Phys. Lett. {\bf B117}, 29 (1982).
\bibitem{kolb2} for a review of baryongenesis please
see \cite{kolb}.
\bibitem{hall} S. Dimopoulos and L. Hall, Phys. Lett. {\bf B196},
135 (1987); M. Claudson, L. Hall, and I. Hinchliffe,
Nucl. Phys. {\bf B241}, 309 (1984).
\bibitem{ellis} G. F. R. Ellis, Class. Quantum Grav. {\bf 5},
891 (1988). 
\bibitem{cs1} T. W. B. Kibble, J. Phys. {\bf A 9},
1387 (1979).
\bibitem{cs2} T. Vachaspati and A. Vilenkin,
Phys. Rev. {\bf D30}, 2036 (1984); N. Turok, Nucl. Phys.
{\bf B242}, 520 (1985); T. W. B. Kibble, Nucl. Phys.
{\bf B252}, 227 (1985); D. P. Bennett, Phys. Rev. {\bf D33},
872 (1986).
\bibitem{cs3} U. Pen, D. N. Spergel and N. Turok,
Phys. Rev. {\bf D49}, 692 (1994).
\bibitem{shavil} Q. Shafi and Z. Vilenkin, Phys. Rev. {\bf D29},
1870 (1984).
\bibitem{hu1} B. L. Hu, J. P. Paz, and Y. H. Zhang,
In Chateau du Pont d'Oye 1992, Proceedings, 227 (1992).
\bibitem{hu2} B. L. Hu and A. Matacz, UMD-PP-94-44,
presented at Workshop on Noise and Order, Los Alamos,
NM, Sept. 1993 (astro-ph/9312012).
\bibitem{allcah} S. M. Allen and J. W. Cahn, Acta metall,
{\bf 27}, 1085 (1979).
\bibitem{bray} A. J. Bray, Adv. Phys. {\bf 43}, 357 (1994).
\bibitem{wilson} K. G. Wilson and J. Kogut, Phys. Rep.
{\bf 12}, 75 (1974); A. Aizenmann, Phys. Rev. Lett.
{\bf 47}, 1 (1981); J. Frohlich, Nucl. Phys. {\bf B200},
[FS4], 281 (1982); M. Luscher and P. Weisz, Nucl. Phys.
{\bf B290} [FS20], 25 (1987).
\bibitem{hol} D. Boyanovsky, H. J. de Vega, and R. Holman,
Phys. Rev. {\bf D49}, 2769 (1994) and references therein.
\bibitem{langer} J. S. Langer, Ann. Phys. {\bf 65}, 53 (1971).
\bibitem{gleiser} M. Gleiser and R. O. Ramos, Phys. Rev. {\bf D50},
2441 (1994); A. Hosoya and M. Sakagami, Phys. Rev. {\bf D29},
2228 (1984); M. Morikawa, Phys. Rev. {\bf D33}, 3607 (1986).
\bibitem{fintemp} B. L. Hu, Phys. Lett. {\bf 108B}, 19 (1982);
Phys. Lett. {\bf 123B}, 189 (1983); I. H. Redmount and F. R. Ruiz,
Phys. Rev. {\bf D39}, 2289 (1989).
\end{references}
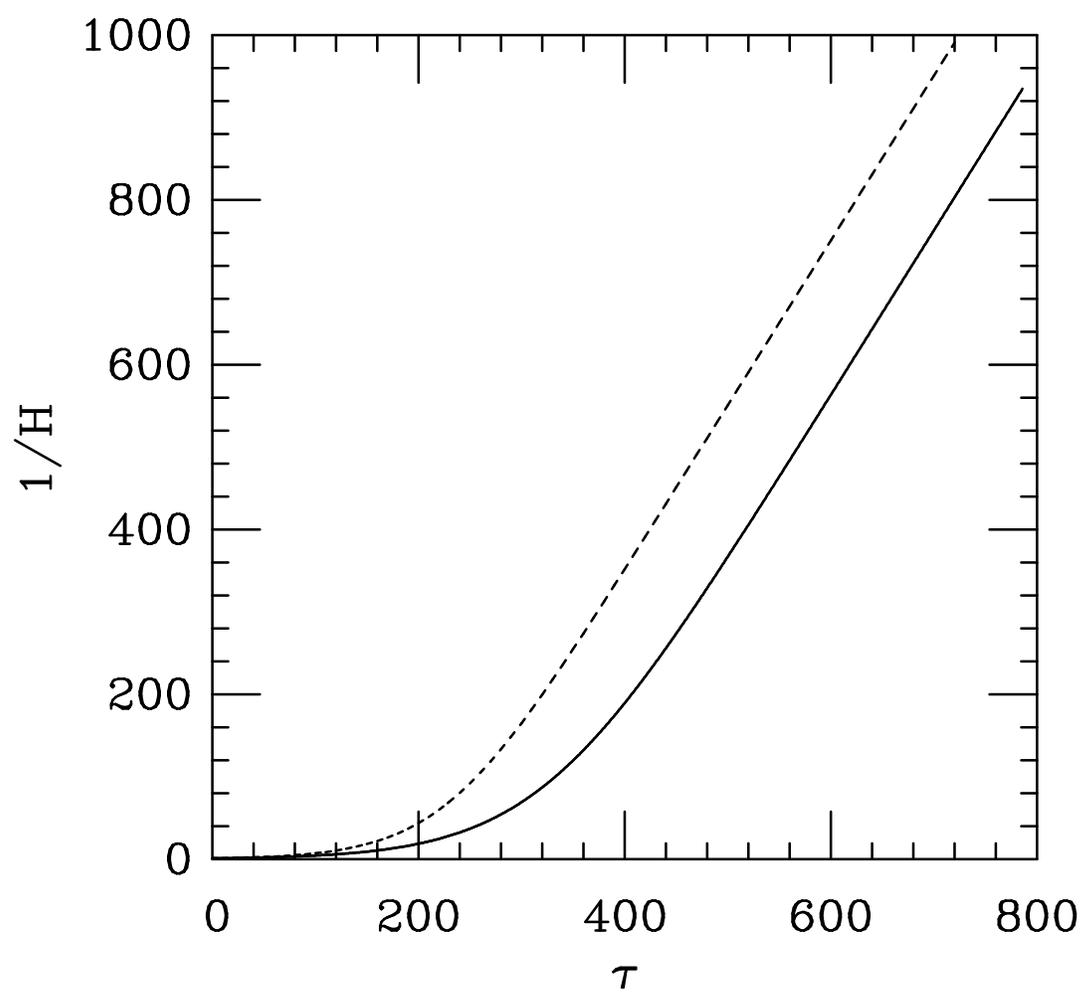
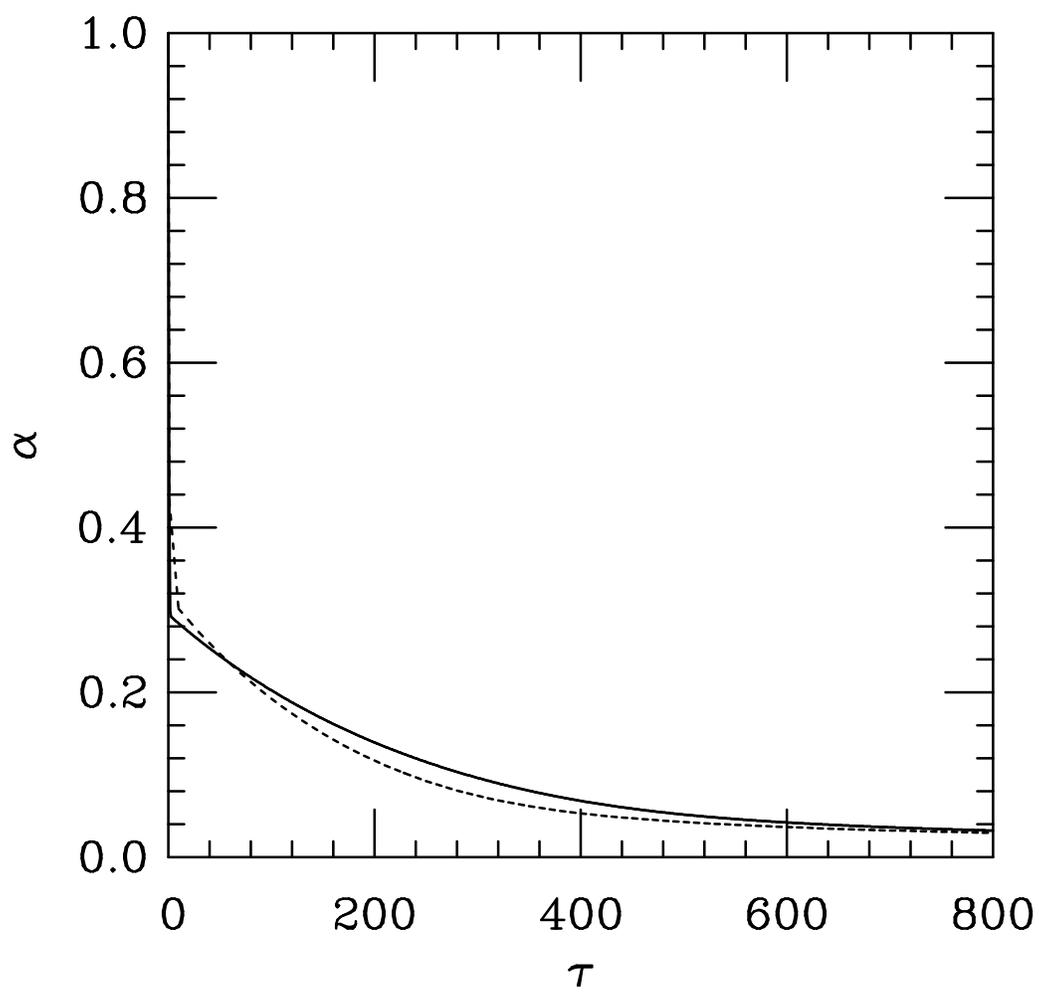
\end{document}